# Microwave Harmonic Emission in $MgB_2$ Superconductor: Comparison with $YBa_2Cu_3O_7$

A. Agliolo Gallitto[1], G. Bonsignore[1], E. Di Gennaro[2], G. Giunchi[3], M. Li Vigni[1] and
P. Manfrinetti[4]

[1]*CNISM and Dipartimento di Scienze Fisiche ed Astronomiche*
*Università di Palermo, Via Archirafi 36, I-90123 Palermo, Italy*

[2]*Dipartimento di Scienze Fisiche Università di Napoli "Federico II"*
*Piazzale Tecchio 80, I-80125 Napoli, Italy*

[3]*EDISON S.p.A., Divisione Ricerca e Sviluppo, Via U. Bassi 2, I-20159 Milano, Italy*

[4]*INFM-LAMIA/CNR, Dipartimento di Chimica e Chimica Industriale*
*Via Dodecaneso 31, I-16146 Genova, Italy*

**Abstract**

We report results of microwave second-harmonic generation in ceramic samples of $MgB_2$, prepared by different methods. The SH signal has been investigated as a function of the temperature and the static magnetic field. The results are discussed in the framework of models reported in the literature. We show that the peculiarities of the SH signal are related to the specific properties of the sample. A comparison with the results obtained in ceramic and crystalline $YBa_2Cu_3O_7$ shows that the second-harmonic emission in $MgB_2$ is weaker than that observed in ceramic $YBa_2Cu_3O_7$.

**Corresponding Author:** Aurelio Agliolo Gallitto
Dipartimento di Scienze Fisiche ed Astronomiche
Università di Palermo
Via Archirafi 36, I-90123 Palermo (Italy)
Tel +39.091.6234207 – Fax +39.091.6162461
agliolo@fisica.unipa.it – www.fisica.unipa.it

## 1. Introduction

It has been widely shown that superconductors (SC) are characterized by reduced microwave (mw) energy losses. For this reason, they are particularly recommended to manufacture components operating at mw frequencies [1]. During long time the most important application of superconductors has concerned the implementation of mw cavities for particle accelerators. Nowadays, they are used in a large variety of mw electronic devices [2,3].

The main factor limiting the application of SC in the implementation of passive mw components is the presence of markedly nonlinear effects, which arise when SC are exposed to intense mw fields [4]. The nonlinear effects manifest through input-power dependence of the mw surface impedance [5, 6], intermodulation distortion [7], generation of signals at the harmonic frequencies of the driving field [8-10]. In order to improve the performance of devices it is important to know the power and temperature ranges in which the nonlinear effects arise. These studies are as well important for understanding the fundamental physics of the superconductivity [4-10].

Soon after the discovery of magnesium diboride in 2001 by Nagamatzu et al. [11], several groups devoted their attention to investigate the properties of such material. Experimental and theoretical studies have shown that $MgB_2$ could be a promising candidate for technological applications [12]. The properties which make this material advantageous for applications are several: the simple composition; the relatively high critical temperature; the large coherence length, which makes the materials less susceptible to structural defects, like grain boundaries, as well as point defects, like impurities; the ductility and malleability of the compound, related to its metallic nature. Because of these promising properties, it is worth to investigate the nonlinear mw response of $MgB_2$ and to compare the results with those obtained in high-$T_c$ superconductors.

In this work we report results of mw second harmonic (SH) generation in two bulk samples of $MgB_2$ prepared by different methods. The results are compared with those obtained in ceramic and crystalline $YBa_2Cu_3O_7$ (YBCO) samples. The SH signal intensity has been investigated as a function of the temperature and DC magnetic field. Particular attention has been devoted to the response at low magnetic fields, where the harmonic emission due to Abrikosov fluxon dynamics is not present. We show that the peculiarities of the SH signal depend on the specific properties of the samples. Furthermore, a comparison between the SH response in ceramic $MgB_2$ and YBCO has shown that the SH emission is less enhanced in $MgB_2$ than in YBCO.

## 2. Samples

Two samples of ceramic $MgB_2$ have been prepared with different methods. Sample B#1 was prepared, using the reactive liquid Mg infiltration technology [13]. Micron size amorphous boron powder (Grade I, 98% purity, Stark H.C., Germany) and pure liquid magnesium are inserted in a stainless steel container, with a weight ratio Mg/B over the stoichiometric value ($\approx 0.55$); the container was sealed by conventional tungsten inert gas welding procedure, with some air trapped inside the B powder; a thermal annealing at 900 °C for 30 min was performed. It has been shown that $MgB_2$ samples produced in this way have a density of 2.40 g/cm$^3$ and consist predominantly of micron size grains, although larger grains, of a few microns in size, are also present [14]. From the final product, we have extracted a sample of approximate dimensions $2 \times 3 \times 0.5$ mm$^3$, in which the largest faces have been mechanically polished to obtain very smooth surfaces. The sample undergoes a superconducting transition with on-set $T_c \approx 38.7$ K and $\Delta T_c \approx 0.7$ K. The mw response of this sample in the linear regime has been previously measured by the hot-finger cavity perturbation technique at $\approx 9$ GHz [15]. From the results of mw surface resistance we have deduced the residual low-temperature resistance and the RRR ratio, obtaining $R_s(T \to 0) < 0.5$ m$\Omega$ and RRR $\equiv \rho(300$ K$)/\rho(T_c) = 4.2$. Sample B#2 has been prepared using the one step method; crystalline B and Mg, put in Ta crucibles welded under argon and closed in quartz tubes under vacuum, were heated up to 950°C [16]. The sample, of approximate dimensions $2 \times 3 \times 0.5$ mm$^3$, undergoes a superconducting transition with on-set $T_c \approx 38.9$ K and $\Delta T_c \approx 0.3$ K.

In order to compare the results obtained in MgB$_2$ with those obtained in high-$T_c$ cuprate superconductors we report results obtained in a single crystal and a ceramic YBCO samples. The YBCO crystal, of approximate dimensions $3 \times 2.5 \times 0.2$ mm$^3$ ($c$-axis parallel to the shortest edge), has $T_c = 92.2$ K and $\Delta T_c \approx 0.1$ K. The bulk ceramic YBCO sample, of cylindrical shape (1.5 mm $\varnothing \times 3$ mm long), has $T_c \approx 89$ K and $\Delta T_c \approx 2$ K.

## 3. Experimental Apparatus

The nonlinear mw response has been investigated by the harmonic generation technique. The intensity of the SH signals has been measured by a nonlinear microwave spectrometer; Fig. 1 shows the block diagram of a spectrometer.

The basic element of the apparatus is a bimodal cavity resonating at the two angular frequencies $\omega$ and $2\omega$. The $\omega$-mode is obtained by inserting a metallic stick into a rectangular copper cavity. The intensity of the magnetic field of the $\omega$-mode varies along the stick as $H(z,\omega) = H_1 \cos(\pi z/2L)\cos(\omega t)$, where $L$ is the length of the stick ($z$-axis parallel to the stick). The harmonic mode is the TE$_{102}$ mode of the rectangular cavity [17], resonating at $\approx 6$ GHz. The distribution of magnetic fields of the fundamental and harmonic modes is shown in Fig. 2. A dielectric stick, inserted in the region of the cavity where is present only the $2\omega$-electrical field, allows properly tuning the SH frequency.

The $\omega$-mode is fed by a train of microwave pulses. A low-power microwave oscillator generates a continuous wave at the desired frequency; a pin-diode, triggered by a pulse generator, allows obtaining a train of microwave pulses with repetition rate ranging from 1 to 200 pps and pulse width ranging from 1 μs to 1 ms; the pulsed field is then amplified up to about 50 W of peak power. A low-pass filter at the input of the cavity cuts any harmonic content by more than 60 dB.

The harmonic signals generated by the sample are filtered by a band-pass filter, with more than 60-dB rejection at the $\omega$-frequency, and are detected by a superheterodyne receiver. The superheterodyne receiver consists of a mixer, a local oscillator at $\omega_L = 2\omega + 30$ MHz and an IF amplifier at 30 MHz with 1 MHz bandwidth. It allows detecting a 30 MHz signal whose intensity is proportional to the harmonic power emitted by the sample. The signal is displayed by an oscilloscope and recorded on a personal computer.

The sample is located in a region of the cavity where the fields $\boldsymbol{H}(\omega)$ and $\boldsymbol{H}(2\omega)$ are maximal and parallel to each other. The cavity is placed between the poles of an electromagnet, which can generate a static field, $H_0$, up to $\approx 10$ kOe. Two additional coils, parallel to the main coils of the magnet and independently fed, are used to reduce the residual field within 0.1 Oe and generate fields of a few oersteds. A LHe cryostat and a temperature controller allow working either at a constant temperature or at temperature varying with a constant rate in the range $4.2 \div 300$ K. The apparatus allows investigating SH signals as a function of temperature, DC magnetic field and input power. In this work, the attention has been devoted to the SH response at external fields smaller than the lower critical field. All the measured signals here reported have been obtained with $\boldsymbol{H}_0 \parallel \boldsymbol{H}(\omega) \parallel \boldsymbol{H}(2\omega)$.

## 4. Experimental results and discussion

Fig. 3 shows the SH signal intensity as a function of the reduced temperature in the samples of ceramic (a) and crystalline (b) YBCO. As one can see, in the crystal the SH emission is essentially observed in a narrow range of temperatures near $T_c$; only a much weaker signal is detected at lower temperatures. On the contrary, the SH emission in the ceramic sample is significant in all the range of temperatures below $T_c$, showing a near-$T_c$ peak as well. These findings, as well as a detailed study of the peculiarities of the low-$T$ signal, have allowed establishing that the low-$T$ and low-field SH emission is ascribable to nonlinear processes occurring in weak links. High quality crystals, in which very few weak links are most likely present, exhibit a very weak SH signal at low temperatures.

The near-$T_c$ peak detected in both samples can be ascribed to the same mechanism; it has been extensively investigated in single crystals where it is not affected by the signal due to the weak links [8, 9]. Measurements performed at temperatures close to $T_c$ at different values of the input power level and DC magnetic field, here not reported, have shown that on increasing $H_0$, and/or $H(\omega)$, the peak broadens and its maximum shifts towards lower temperatures. Moreover, the input power dependence of the SH signal is less than quadratic. All these peculiarities have been quite well accounted for by a phenomenological theory, based on the two-fluid model, where the additional hypothesis is made that the mw and the DC magnetic fields, penetrating the surface layers of the sample, weakly perturb the partial concentrations of the normal and condensed electrons [8, 9].

Fig. 4 shows the field dependence of the SH signal for the YBCO samples, ceramic (a) and crystalline (b). As expected from symmetry considerations, the SH signal is zero at $H_0 = 0$; on increasing $H_0$, the SH signal intensity rapidly increases and shows a maximum at few Oe; it has been shown that the position of the maximum depends on the input power level. As one can see, though in both samples the SH vs. $H_0$ curves has qualitatively the same behaviour, they strongly differ for the intensity, which in the ceramic sample is three order of magnitude higher. We would remark that after the samples were exposed to DC magnetic fields of the order of 100 Oe the low-field structure disappears irreversibly; this finding confirms that the low-field SH signal is due to processes occurring in weak links, indeed, when the samples go in the mixed state, weak links are decoupled by the high magnetic field and/or the trapped flux. The peculiarities of the low-field SH signal in ceramic YBCO samples have been accounted for by assuming that supercurrents are induced by the $H_0$ and $H(\omega)$ fields in loops containing Josephson junctions; so, the harmonic emission is strictly related to the intrinsic nonlinearity of the Josephson current [10,18].

Fig. 5 shows the SH signal intensity as a function of the temperature for the two $MgB_2$ samples, B#2 (a) and B#1 (b), at $H_0 = 10$ Oe. In sample B#2 the SH emission is noticeable in the whole range of temperatures investigated and exhibits an enhanced peak at temperatures near $T_c$. On the contrary, within our experimental sensibility, no detectable SH emission has been observed in B#1 at low temperatures; the SH emission of this sample is noticeable only in a restrict range of temperatures near $T_c$. A comparison between the two curves shows that, though B#2 exhibits a superconducting transition narrower than B#1, the near-$T_c$ peak is wider in B#2 than in B#1 sample; we think that this occurs because of the influence of the low-$T$ signal. According to what previously discussed, the different behaviour of the SH emission in the two samples of $MgB_2$ can be ascribed to a reduced number of weak links in sample B#1. In any cases, we can note that the low-$T$ SH emission in $MgB_2$ is markedly reduced with respect to that observed in ceramic YBCO; this confirms that in $MgB_2$ only a small amount of grain boundaries acts as weak links, in contrast to what occurs in cuprate superconductors.

Concerning the near-$T_c$ peak, we suggest that also in $MgB_2$ the SH emission is due to modulation of the partial concentrations of the normal and superconducting fluids induced by the mw field. However, the peak width in YBCO is about $T_c/30$, while in MgB$_2$ it is about $T_c/10$; according to what reported in the literature [19], this finding may be ascribed to a different temperature dependence of the densities of the two fluids.

In order to corroborate the idea that, also in MgB$_2$, the low-$T$ and low-field SH emission is due to processes in weak links, we have investigated the field dependence of the SH signal. In Fig. 6 we report the SH-signal intensity as a function of $H_0$ for the MgB$_2$ B#2 sample, at $T = 4.2$ K, obtained by cycling the field between $\pm H_0^{max}$. Plots (a) and (b) differ for the value of $H_0^{max}$. As one can see, sweeping $H_0$ in the range $\pm 5$ Oe, the results are reproduced on increasing and decreasing the field; after $H_0^{max}$ reaches a threshold value, the signal shows a hysteretic behaviour. We would remark that, similarly to what occurs in YBCO, after the sample is exposed to magnetic field of the order of 100 Oe the low-field structure vanishes irreversibly. This corroborates that the mechanism responsible for the low-$T$ and low-field SH emission is the same in YBCO and MgB$_2$. The presence of the magnetic hysteresis, together with the observation that on decreasing the field, after the first run, the signal is observed even for $H_0 = 0$, suggests that trapped flux is present in the sample. On the other hand, the hysteresis is observed at applied fields so weak to rule out the presence of Abrikosov fluxons; so, we think that the harmonic emission involves trapped flux in weak links [20]. It is worth noting that the SH vs. $H_0$ curves of Fig. 4 do not show any hysteresis; however, also in YBCO can be observed magnetic hysteresis provided that the field is spanned in a wider range ($H_0^{max} \approx 50$ Oe).

## 5. Conclusion

We have discussed on microwave second-harmonic response of ceramic MgB$_2$, making a comparison with that of YBCO superconductors. The SH signal has been investigated as function of the temperature and the DC magnetic field. The results show that, similar to what occurs in other high-$T_c$ superconductors, different mechanisms are responsible for the nonlinear microwave response of MgB$_2$; their effectiveness depends on the temperature. The results obtained at low temperatures have shown that, although the presence of weak links in MgB$_2$ does not noticeably affect the transport properties, it is source of nonlinearity at low magnetic fields and low temperatures. At temperatures close to $T_c$, a peak in the temperature dependence of the SH signal is present; it may arise from modulation of the order parameter by the microwave field. The near-$T_c$ peak detected in MgB$_2$ is wider than that observed in YBCO crystals; this finding may be ascribed to a different temperature dependence of the normal and superconducting fluid densities in the two compounds. We would remark that similar results have been obtained in other bulk MgB$_2$ samples, prepared by different methods [21]. In any cases, the harmonic emission in MgB$_2$ is noticeably reduced with respect to that observed in ceramic cuprate superconductors. This finding is of relevance for using MgB$_2$ in the implementation of passive mw devices.

**Figure captions**

Fig. 1 – Block diagram of the nonlinear spectrometer operating at 3–6 GHz.

Fig. 2 – Bimodal cavity: continuous lines show the distribution of the magnetic field of the $\omega$-mode; dashed lines that of the $2\omega$-mode.

Fig. 3 – SH signal intensity as a function of the reduced temperature for YBCO superconductor; (a) ceramic sample at input peak power $\approx$ 36 dBm; (b) single crystal at input peak power ~ 40 dBm.

Fig. 4 – Low-field dependency of the SH signal for YBCO superconductor; (a) ceramic sample at input peak power $\approx$ 36 dBm; (b) single crystal at input peak power ~ 40 dBm.

Fig. 5 – SH signal intensity as a function of the reduced temperature for the two $MgB_2$ samples: B#2 (a) and B#1 (b); $H_0$ = 10 Oe; input peak power $\approx$ 30 dBm.

Fig. 6 – SH signal intensity as a function of the DC magnetic field for the $MgB_2$ B#2 sample, obtained in the ZFC sample on increasing and decreasing the field; $T$ = 4.2 K; input peak power $\approx$ 36 dBm.

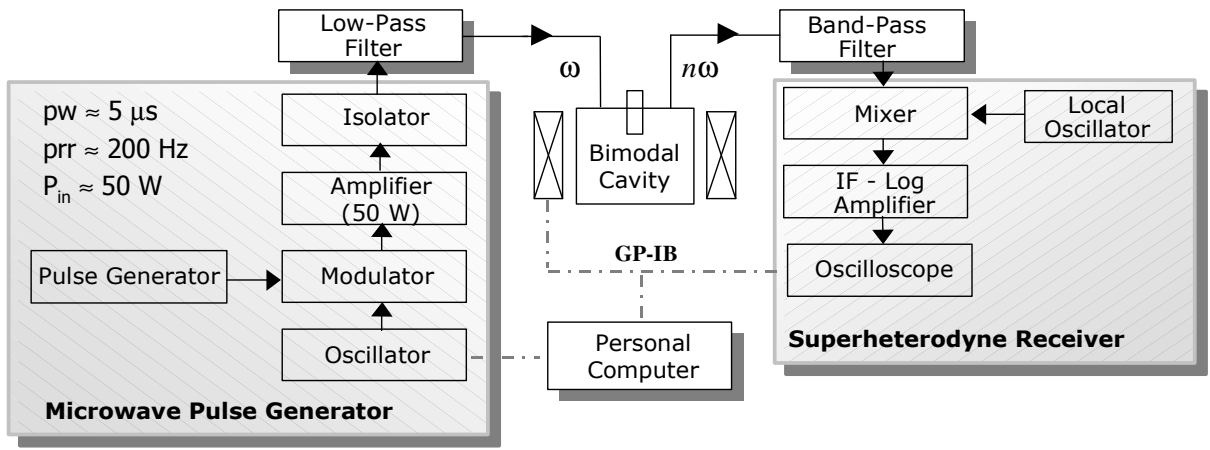

Figure 1

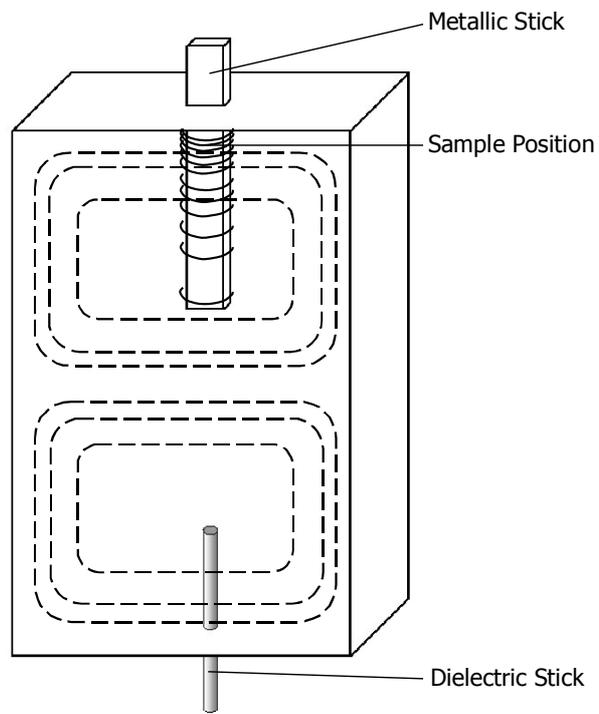

Metallic Stick

Sample Position

Dielectric Stick

Figure 2

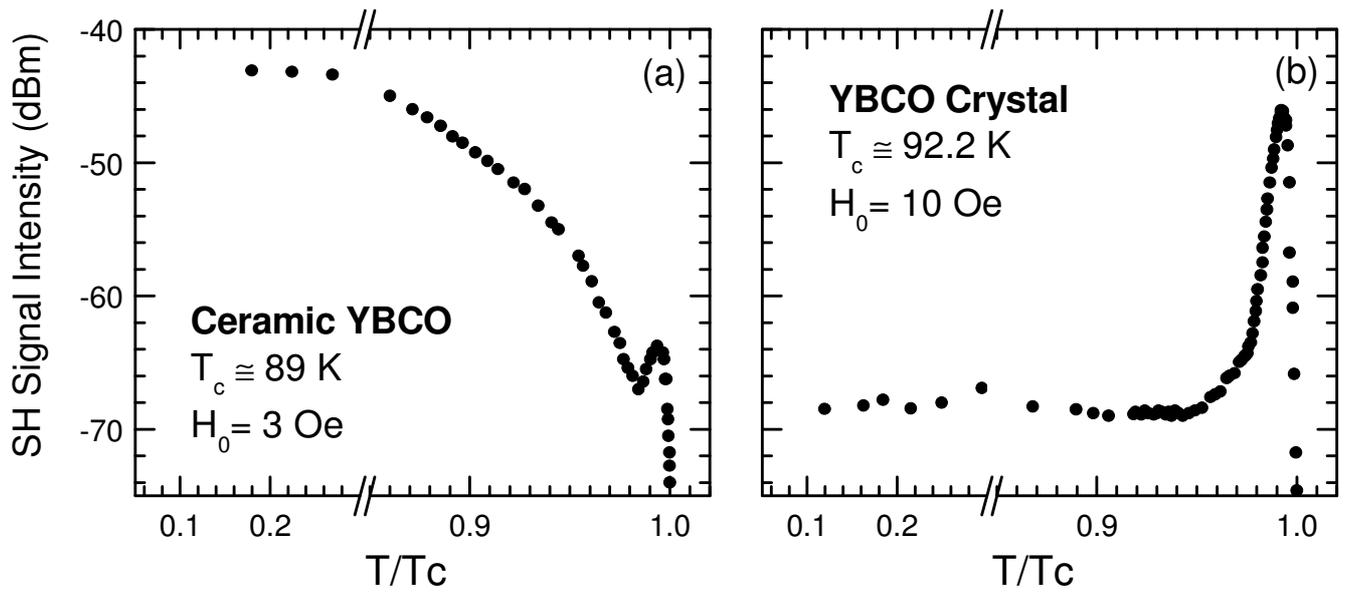

Figure 3

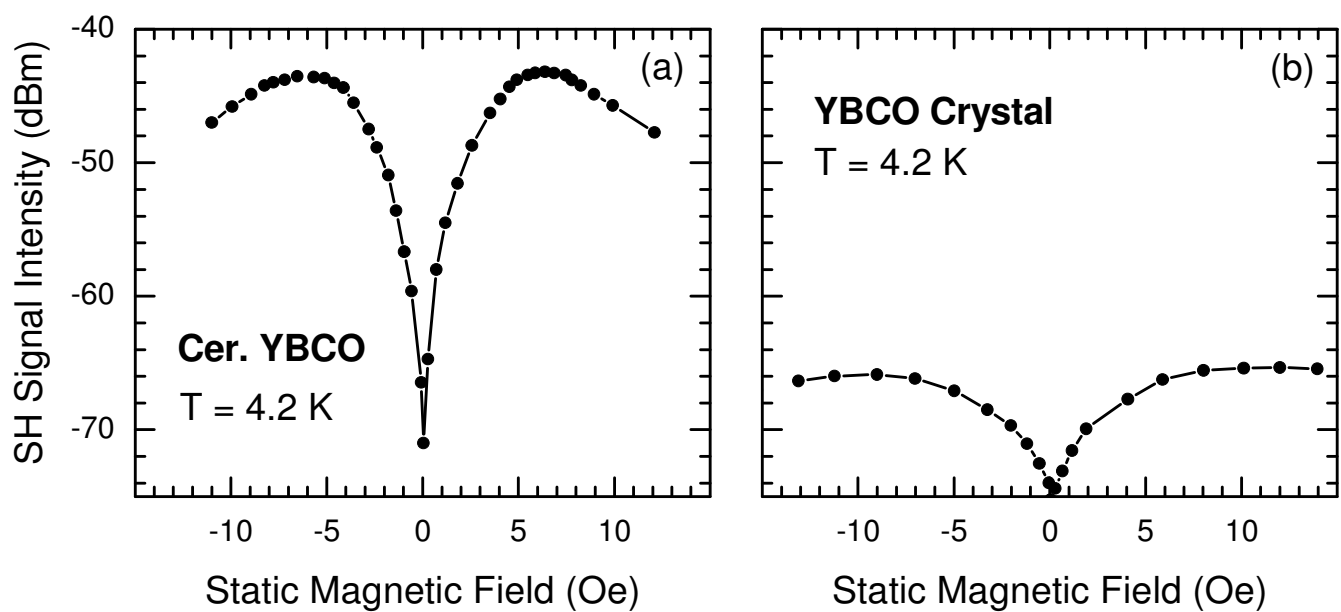

Figure 4

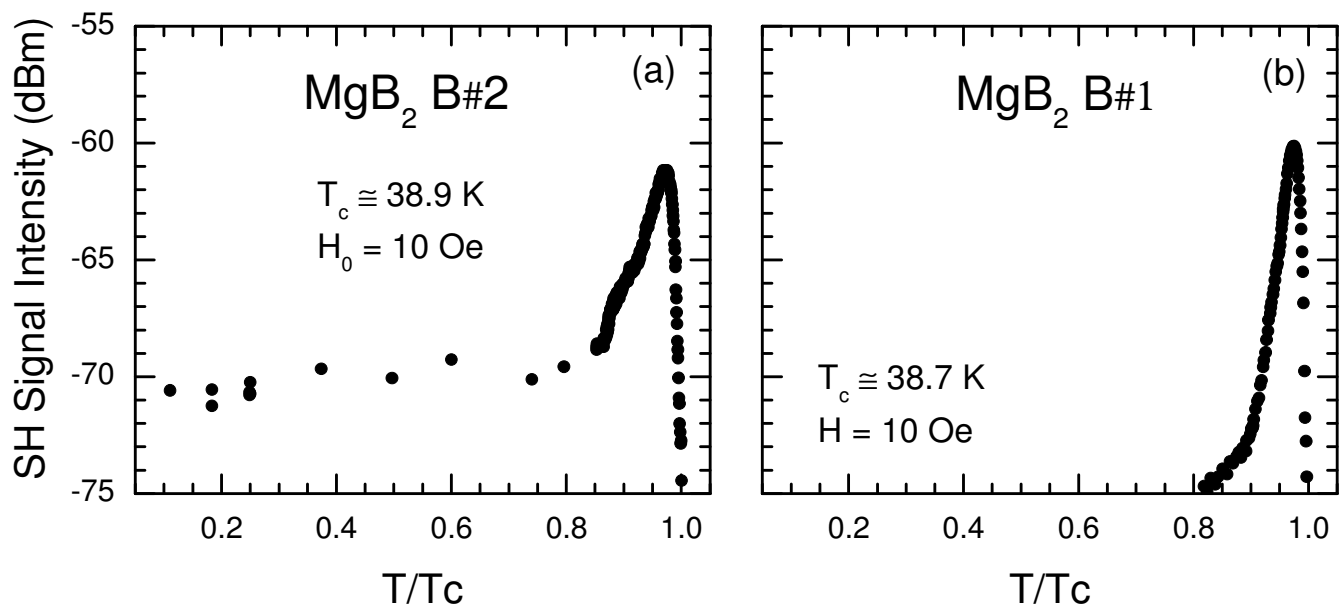

Figure 5

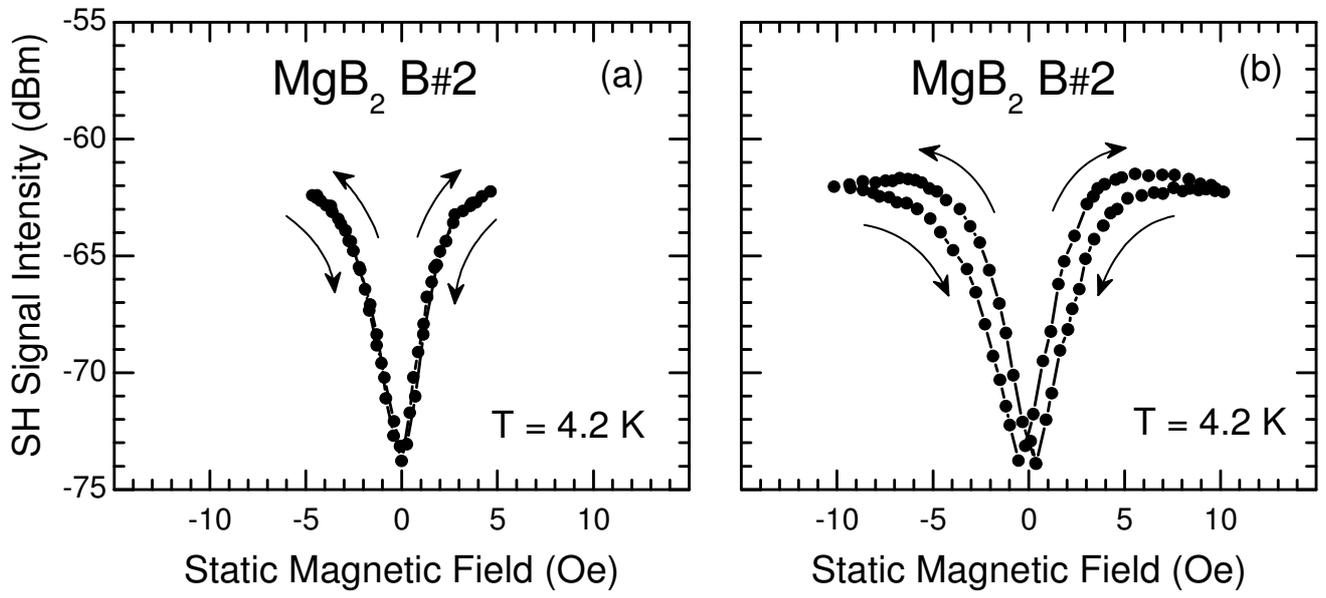

Figure 6